\documentstyle[12pt]{article}

\input epsf.sty

\newcommand{\g}{{\sl g}}

\newcommand{\ft}[2]{{\textstyle\frac{#1}{#2}}}

\newcommand{\Sym}{\mathop{\mbox{\large\bf S}}}

\begin{document}

\begin{titlepage}

\centerline{\large \bf Resummation of target mass corrections
in two-photon processes:}
\centerline{\large \bf twist-two sector.}

\vspace{15mm}

\centerline{\bf A.V. Belitsky$^a$, D. M\"uller$^b$}
\vspace{5mm}

\centerline{\it $^a$C.N.\ Yang Institute for Theoretical Physics}
\centerline{\it State University of New York at Stony Brook}
\centerline{\it NY 11794-3840, Stony Brook, USA}

\vspace{5mm}

\centerline{\it $^b$Fachbereich Physik,
              Universit\"at  Wuppertal}
\centerline{\it D-42097 Wuppertal, Germany}

\vspace{15mm}

\centerline{\bf Abstract}

\vspace{1cm}

We develop a formalism for the resummation of target mass corrections in
off-forward two-photon amplitudes given by a chronological product of
electromagnetic currents, arising in e.g.\ deeply virtual Compton scattering.
The method is based on a relation of composite operators with a definite
twist to harmonic tensors, which form an irreducible representation of the
Lorentz group. We give an application of the framework for the matrix
elements of twist-two operators.

\vspace{6cm}


\vspace{0.5cm}


\end{titlepage}

\section{Hadron mass corrections.}

The leading twist approximation to hard processes in QCD is affected by a
number of multiplicative and additive corrections. To the first class one
obviously attributes radiative corrections in the strong coupling, while
the latter encodes higher twist contributions, which provide power suppressed
corrections. In the case when a given large scale $Q$, that controls the
factorization of a reaction, becomes rather low, one has to take care of
power suppressed effects since they can significantly modify the scaling
behaviour and the magnitude of the leading twist prediction for the
corresponding cross section. Power corrections can be divided into two
classes according to their origin: dynamical and kinematical ones. The
first one comprises of multiparton correlations inside hadrons \cite{Pol80}
and gives rise to new non-perturbative functions. The second one arises
from a separation of composite operators into parts that have definite
transformation properties w.r.t.\ the Lorentz group and thus possess a
well defined geometrical twist, i.e.\ dimension minus spin. This
decomposition provides the so-called Wandzura-Wilczek type contributions
\cite{WanWil77} arising from components having definite symmetry properties
as well as target mass corrections \cite{GeoPol76} stemming from the
subtraction of trace terms in the afore mentioned operators.

In the present paper we develop a formalism for the resummation of the
target mass corrections for off-forward two-photon processes whose amplitudes
are given by a chronological product of electromagnetic currents. This method
is indispensable for a study of hadron mass effects in the cross section of
e.g.\ deeply virtual Compton scattering \cite{MulRobGeyDitHor94,Ji96,Rad96},
--- the processes sensitive to the generalized parton distributions (GPDs).
The present day facilities can offer the $Q^2$ of the order of a few (2-4)
$\mbox{GeV}^2$ (for the reactions in question) and one has to account for
power suppressed contributions to the twist-two and -three observables
arising from the ratio of the nucleon mass to the hard momentum transfer.
Unfortunately, the formalism for resummation of target mass corrections
invented by Georgi and Politzer \cite{GeoPol76} for the deep inelastic
scattering (see also \cite{Nac73} for a first discussion of the topic), and
accepted in all consequent generalizations and applications to forward
electroweak scattering amplitudes, is not directly applicable to off-forward
processes in the context of GPDs. The reason for that is the appearance of
towers of new Lorentz structures in the matrix elements of local operators,
arisen due to the non-zero $t$-channel momentum transfer, which cannot
be handled in the fashion proposed in \cite{GeoPol76}. Note that a
discussion of mass effects in exclusive processes with a much simpler
kinematics can be found in Ref.\ \cite{BalBra99}.

In the present context, instead of dealing with single variable functions,
e.g.\ GPDs, we use spectral functions, the so-called double distributions
(DDs) \cite{MulRobGeyDitHor94,Rad96}, which depend on two variables. This
allows to resum the mass corrections in a straightforward manner and to
derive all-order results in a compact form. After this is accomplished
one can in principle use an inverse integral representation to express DDs
by means of GPDs, however, this transformation requires the knowledge of
the support of GPDs including an unphysical region. Therefore, this
transformation can be hardly used for a successful phenomenological
application.

The consequent presentation runs as follows. In the next section we give
a short introduction to a procedure of the trace subtraction by means of
harmonic polynomials. In Section \ref{Sec-Res} we resum the target mass
effects resulting from the trace subtraction in matrix elements of twist-two
operators for zero and spin-$\ft12$ targets. Then it is shown that the
expansion of these results in powers of $\left( M^2/q^2 \right)^j$ can be
cast in a conventional form of GPDs and be used for numerical predictions
of target mass effects in physical cross sections. Finally, we conclude.

\section{Harmonic polynomials and twist decomposition.}
\label{Sec-twDec}

The goal of our study is an evaluation of the off-forward Compton amplitude
keeping target mass effects stemming from the twist-two contributions in the
light-cone expansion of the off-forward matrix element of the chronological
product $\widetilde T_{\mu\nu} (\ft{x}{2}, - \ft{x}{2}) \equiv \langle P_2 |
T \left\{j_\mu (\ft{x}{2}) j_\nu (- \ft{x}{2}) \right\} | P_1 \rangle$ of
electromagnetic currents $j_\mu = \bar\psi \gamma_\mu \psi$. Since these are
entirely kinematical effects, we consider dynamical higher twist effects
given by multiparticle operators as independent and, therefore, we safely
neglect them. Note, however, that the trace subtraction in twist-three
contributions in the Wandzura-Wilczek approximation will generate kinematical
mass effects accompanied by leading twist operators. We will comment on this
issue in the concluding section. In the approximation, we are considering,
the Fourier transformed two-photon amplitude reads
\begin{eqnarray}
\label{AmplitudeT}
T_{\mu \nu}
\!\!\!&=&\!\!\!
\int d^4 x \ {\rm e}^{i x \cdot q} \ \widetilde T_{\mu\nu}
\left( \ft{x}{2}, - \ft{x}{2} \right)
=
\frac{1}{\pi^{2}}
\int d^4 x \ \frac{x_\sigma}{[- x^2 + i 0]^{2}}
\nonumber\\
&\times&\!\!\! \langle P_2 |
\cos \left( x \cdot q \right) S_{\mu \nu; \rho \sigma}
{^V\!\!{\cal O}_\rho} \! \left( \ft{x}{2}, - \ft{x}{2} \right)
+ \sin \left( x \cdot q \right) \epsilon_{\mu \nu \rho \sigma}
{^A {\cal O}_\rho} \! \left( \ft{x}{2}, - \ft{x}{2} \right)
| P_1 \rangle ,
\end{eqnarray}
where $S_{\mu\nu;\rho\sigma} = g_{\mu\rho} g_{\nu\sigma} + g_{\mu\sigma}
g_{\nu\rho} - g_{\mu\nu} g_{\rho\sigma}$ and for the totally
antisymmetric tensor we use the normalization $\epsilon^{0123} = 1$.
The vector $q$ is a semi-sum of incoming $q_{1 \nu}$ and outgoing
$q_{2 \mu}$ photon momenta, $q = \ft12 (q_1 + q_2)$. The non-local
operators in Eq.\ (\ref{AmplitudeT}) are ${^{\mit\Gamma}\! {\cal O}_\rho}
(x, -x) = \bar\psi (-x) [-x, x] {\mit\Gamma}_\rho \psi (x)$ with a Dirac
matrix ${\mit\Gamma}_\rho$ which equals $\gamma_\rho$ or $\gamma_\rho
\gamma_5$ for vector ($V$) or axial-vector ($A$) sectors, respectively.
The path ordered exponential $[-x, x]$ will be dropped everywhere in the
following formulae.

We are aiming now in a decomposition of ${\cal O}_{\rho}$ into traceless
operators ${\cal R}^2_\rho$ and ${\cal R}^3_\rho$ with definite geometrical
twist-two and -three, respectively:
\begin{equation}
{\cal O}_\rho \! \left( x, - x \right)
=
{\cal R}^2_\rho \! \left( x, - x \right)
+
{\cal R}^3_\rho \! \left( x, - x \right)
+
{\cal R}^r_\rho \! \left( x, - x \right) .
\end{equation}
To this end, we expand first the non-local operator ${\cal O}_\rho \!
\left( x, - x \right)$ in a Taylor series
\begin{eqnarray}
\label{Def-Ope}
{\cal O}_\rho (x, -x)
= \sum_{j = 0}^{\infty} \frac{(-i)^j}{j!}
x_{\mu_1} \dots x_{\mu_j} {\cal O}_{\rho; \mu_1 \dots \mu_j} ,
\qquad\mbox{with}\qquad
{\cal O}_{\rho; \mu_1 \dots \mu_j} =
\bar\psi {\mit\Gamma}_\rho \,
i\!\stackrel{\leftrightarrow}{\cal D}_{\mu_1}
\dots
i\!\stackrel{\leftrightarrow}{\cal D}_{\mu_j}
\psi ,
\end{eqnarray}
where $\stackrel{\leftrightarrow}{\cal D}_{\mu} \ = \
\stackrel{\rightarrow}{\cal D}_{\mu} - \stackrel{\leftarrow}{\cal D}_{\mu}$
with the covariant derivative ${\cal D}_\mu = \partial_\mu - i \g B_\mu$,
and then we perform the twist decomposition for the local operators ${\cal
O}_{\rho;j} = x_{\mu_1} \dots x_{\mu_j} {\cal O}_{\rho; \mu_1 \dots \mu_j}$.
An effective way of trace subtraction is to solve the sufficient condition
$\partial^2 H^{j} \left( x^2, \partial^2 \right) T_j(x) = 0$
\cite{Vil68,BalBra89,RobGeyLaz99}, where $T_j$ results from the contraction
of a rank-$j$ tensor with the product of four-vectors $x_{\mu_1} \dots
x_{\mu_j}$, and $H^j$ is a subtraction operator. The solution is given by
the so-called harmonic polynomials \cite{Vil68}
\begin{eqnarray}
H^{j} \left( x^2, \partial^2 \right) = \sum_{k=0}^{[j/2]}
\frac{\Gamma\left(j - k + 1 \right)}{k!\Gamma\left( j + 1 \right)}
\left(-\frac{x^2}{4}\right)^k \left(\partial^2 \right)^k .
\end{eqnarray}
For operators having a free vector index one has an extra condition so that
in total the tracelessness implies $\partial^2 {\cal R}^i_{\rho; j} = 0$
and $\partial_\rho {\cal R}^i_{\rho; j} = 0$, for $i = 2, 3$, where the
twist-two operators are completely symmetric while the twist-three operators
are antisymmetric in a pair of their Lorentz indices. Employing the equation
of motion for the massless Dirac field, one can show that ${\cal R}^r$ is
proportional to total derivative, $\nabla$, squared, apart from
multiparticle operators, ${\cal R}^r_{\rho} \propto - \nabla^2
{\cal R}^2_{\rho} + \, {\rm multiparticle \ operators}$. When it is
sandwiched between hadronic states, it generates contributions proportional
either to the $t$-channel momentum transfer squared or given by multi-parton
correlations functions. Both of them will be neglected in the following.
Thus, our considerations will be limited by the condition $M^2 \gg \Delta^2$,
with $M^2 = P_1^2 = P_2^2$ and $\Delta^2$, being the hadron mass and
$t$-channel momentum transfer, respectively.

A straightforward algebra leads to the desired decomposition in operators
with definite geometrical twist:
\begin{equation}
{\cal O}_{\rho; j} =  {\cal R}^2_{\rho; j}
+ \frac{2 j}{j + 1} {\cal R}^3_{\rho; j} + {\cal R}^r_{\rho; j} ,
\end{equation}
where the traceless twist-two and -three operators read (cf.\
\cite{RobGeyLaz99})
\begin{eqnarray}
\label{R2traceless}
{\cal R}^2_{\rho; j}\!\!\!&=&\!\!\!
\frac{1}{j + 1}\partial_\rho H^{j + 1} \left( x^2, \partial^2 \right)
x_\sigma {\cal O}_{\sigma; j} ,
\qquad\qquad\qquad\qquad\qquad\qquad\quad\
j \geq 0 ,\\
\label{R3traceless}
{\cal R}^3_{\rho; j} \!\!\!&=&\!\!\! \frac{1}{2 j}
\left[ g_{\rho\sigma} x \cdot \partial - x_\sigma \partial_\rho \right]
\left[ g_{\sigma\tau}  - \frac{1}{j + 1} x_{\sigma} \partial_\tau \right]
H^{j} \left( x^2, \partial^2 \right) {\cal O}_{\tau; j} ,
\qquad
j \geq 1 ,
\end{eqnarray}
and the remainder is
\begin{eqnarray*}
{\cal R}^r_{\rho; j} \!\!\!&=&\!\!\! \frac{1}{j + 1}
\Bigg\{ \partial_\rho
\left[ 1 - H^{j + 1} \left( x^2, \partial^2 \right) \right] x_\sigma
+ \left[ g_{\rho\sigma} x \cdot \partial - x_\sigma \partial_\rho \right]
\left[ 1 - H^{j} \left( x^2, \partial^2 \right) \right]
\nonumber\\
&&\qquad + \frac{1}{j + 1}
\left[ x_{\rho} x \cdot \partial - x^2 \partial_\rho \right] \partial_\sigma
H^{j} \left( x^2, \partial^2 \right)
\Bigg\} {\cal O}_{\sigma; j} .
\end{eqnarray*}
It is obvious that the twist-two operators (\ref{R2traceless}) are
traceless. Using $\partial^2 H^j \left( x^2, \partial^2 \right) {\cal
O}_{\mu; j} = 0$ and Euler theorem $x \cdot \partial \ \partial_\mu H^j
\left( x^2, \partial^2 \right) {\cal O}_{\mu; j} = (j - 1) \partial_\mu H^j
\left( x^2, \partial^2 \right) {\cal O}_{\mu; j}$, it is an easy task to
check that the operators (\ref{R3traceless}) are traceless too.

\section{Resummation within DD formalism.}
\label{Sec-Res}

In this section we will give a general framework for the resummation of
kinematical mass corrections. For simplicity we consider here only the
twist-two sector. The major steps consist of: a parametrization of the
off-forward matrix elements of symmetric local operators by means of
moments of DDs, the trace subtraction with the harmonic projectors $H^j$,
then a Fourier transformation to the momentum space and, finally, a
resummation of infinite series.

In order to demonstrate the main features of our formalism, let us consider
first matrix elements of twist-two operators sandwiched between states of
spinless target $\langle P_2| {\cal R}^2_{\rho; j} |P_1\rangle$. Here only
the matrix elements of the parity even sector are relevant and read
\begin{equation}
\label{Def-RedMattw2}
\langle P_2|
{^V\!\!{\cal O}^{\rm sym}_{\rho; \mu_1 \cdots \mu_j}}
|P_1\rangle
= P_{\{\rho} P_{\mu_1} \cdots P_{\mu_j\}} B_{j + 1, j + 1}
+ \cdots
+ \Delta_{\{\rho} \Delta_{\mu_1} \cdots \Delta_{\mu_j\}} B_{j + 1, 0}
+ \dots \ .
\end{equation}
They are build from the two vectors $P = P_1 + P_2$ and $\Delta = P_2 - P_1$,
$\{ \cdots \}$ denotes symmetrization of corresponding indices and the
ellipsis stands for possible terms containing the metric tensor. Since we
act on this expression with $H^j$ we can neglect the latter. After
projection with harmonic polynomials via Eq.\ (\ref{R2traceless}) one gets
\begin{eqnarray}
\label{FinSumTw2}
\langle P_2| {^V\!{\cal R}^2_{\rho;j}} |P_1\rangle
\!\!\!&=&\!\!\!
\frac{1}{j + 1} \partial_\rho H^{j + 1} \left( x^2, \partial^2 \right)
\sum_{k=0}^{j+1} \left( x \cdot P \right)^{j + 1 - k}
\left( x \cdot \Delta \right)^{k} B_{j + 1, j + 1 - k} \\
&=&\!\!\! x_{\mu_1} \dots x_{\mu_j}
\Sym_{\rho \mu_1 \dots \mu_j}
\left\{
P_\rho P_{\mu_1} \dots P_{\mu_j} B_{j+1, j+1}
+ \cdots
+ \Delta_\rho \Delta_{\mu_1} \dots \Delta_{\mu_j} B_{j+1, 0}
\right\} ,
\nonumber
\end{eqnarray}
with the operation $\Sym$ standing for symmetrization and trace subtraction.
The first line gives a compact expression for subtracted operator which is
extremely convenient for consequent considerations and is the basis of our
formalism.

On the other hand we can deduce the twist-two operator from Eq.\
(\ref{Def-RedMattw2}) by a contraction of all Lorentz indices with a
light-like vector $n$ so that the trace terms vanish identically. This
tells us that the coefficients $B_{j + 1, j + 1 - k}$ in front of the
Lorentz tensors are related to the reduced matrix elements of light-cone
operators and can be represented in terms of moments of conventional
leading twist GPDs or DDs via
\begin{equation}
\label{BversusDD}
B_{j, j - k} = \frac{1}{k!}
\left. \frac{\partial^k}{\partial\eta^k} \right|_{\eta = 0}
\int_{-1}^{1} dx\; x^{j - 1} B (x,\eta)
= \left({j \atop k}\right) \int_{{\mit\Omega}}
dy \, dz \,  y^{j - k} z^{k} f (y, z)
\end{equation}
where $0 \le k \le j$, $1 \le j$ and the integration domain is ${\mit\Omega}
= \{ - 1 \leq y \leq 1 , - 1 + |y| \leq z \leq 1 - |y| \}$. Consequently,
the parametrization\footnote{This is not the definition given in Refs.\
\cite{MulRobGeyDitHor94,Rad96} which suffered from inconsistencies due to
omitted Lorentz structures as noted in \cite{PolWei99}. Our solution of
this problem differs also from \cite{PolWei99} where a two-component
parametrization has been suggested.} of the matrix element for the non-local
symmetric operator and the light-ray operators coincide up to terms
proportional to $x_\rho$ or $x^2$:
\begin{equation}
\label{definitionDD}
\langle P_2 |
{^V\!\!{\cal O}^{\rm sym}_\rho (x, - x)}
| P_1 \rangle
= \int_{{\mit\Omega}} dy \, dz \, f (y, z) {\cal P}_\rho
\ {\rm e}^{- i x \cdot {\cal P}} + \dots,
\end{equation}
where we introduced a new shorthand notation ${\cal P}_\mu
\equiv y P_\mu + z \Delta_\mu$.

The resummation of matrix elements of traceless local operators can be
performed in the formalism of DD $f (y, z)$. By means of Eq.\
(\ref{BversusDD}) it is easy to see that the finite sum w.r.t.\ $k$ in
Eq.\ (\ref{FinSumTw2}) provides
\begin{equation}
\label{LocalOperDD}
\langle P_2| {^V\!{\cal R}^2_{\rho;j}} |P_1\rangle
= \int_{{\mit\Omega}} dy \, dz\, f (y, z)
\frac{1}{j+1} \partial_\rho H^{j + 1} \left( x^2, \partial^2 \right)
\left( x \cdot {\cal P} \right)^{j + 1} .
\end{equation}
The evaluation of projection of harmonic polynomials results into
conventional Chebyshev polynomials related to the irreducible
representation of the orthogonal group $SO (4)$
\begin{equation}
\label{ChebyshevResumm}
H^{j} \left( x^2, \partial^2 \right)
\left( x \cdot {\cal P} \right)^{j}
= \left( \frac{x^2 {\cal P}^2}{4} \right)^{j/2}
U_{j} \left( \frac{x \cdot {\cal P}}{\sqrt{x^2 {\cal P}^2}} \right) .
\end{equation}
This result when plugged into Eq.\ (\ref{LocalOperDD}) gives our final
expression for the twist-two traceless local operators. We will use it
after we perform the Fourier transformation in Eq.\ (\ref{AmplitudeT}).
We accomplish it making use of the following general formula
\begin{equation}
\int d^4 x {\rm e}^{i x \cdot q}
\frac{x_{\mu_1} \dots x_{\mu_j}}{[- x^2]^k}
= (- i)^{j + 1} 2^{4 - 2 k + j} \pi^{2}
\frac{{\mit\Gamma} (2 - k + j)}{{\mit\Gamma} (k)}
\frac{q_{\mu_1} \dots q_{\mu_j}}{[- q^2]^{2 - k + j}} ,
\end{equation}
where on the r.h.s.\ we dropped all terms involving the metric tensor
$g_{\mu_i \mu_k}$, since they vanish when contracted with a traceless
tensor. Thus, we get
\begin{eqnarray}
\label{Integral}
&&\int d^4 x {\rm e}^{i x \cdot q} \frac{x_\sigma}{[- x^2]^{2}}
\frac{x_{\mu_1}}{2} \dots \frac{x_{\mu_j}}{2}
\langle P_2 | {^V\!{\cal R}^2_{\rho; \mu_1 \dots \mu_j}} | P_1 \rangle
\nonumber\\
&&\qquad\qquad\qquad\qquad\qquad\qquad= i^{j + 2} \pi^{2}
{\mit\Gamma} (j + 1)
\frac{1}{q^2} {\mit\Pi}_{\sigma \mu_1}
\frac{q_{\mu_2}}{q^2} \dots \frac{q_{\mu_j}}{q^2}
\langle P_2 | {^V\!{\cal R}^2_{\rho; \mu_1 \dots \mu_j}} | P_1 \rangle ,
\end{eqnarray}
with projector
\begin{equation}
{\mit\Pi}_{\mu\nu}
\equiv g_{\mu\nu} - \frac{2}{q^2} q_\mu q_\nu ,
\qquad
{\mit\Pi}_{\mu\rho} {\mit\Pi}_{\rho\nu} = g_{\mu\nu} .
\end{equation}
Then the Fourier transformation of the twist-two contribution to the Compton
amplitude gives
\begin{eqnarray}
\label{Fourier}
&&\int d^4 x {\rm e}^{i x \cdot q} \frac{x_\sigma}{[- x^2]^2}
\langle P_2 |
{^V\!{\cal R}^2_\rho} \left( \ft{x}{2}, - \ft{x}{2} \right)
| P_1 \rangle \nonumber\\
&&\qquad\qquad\qquad\qquad\qquad
= 2 \pi^2
\left\{
\frac{q_\sigma}{q^2} \sum_{j = 0}^{\infty}
\langle P_2 | {^V\!\tilde {\cal R}^2_{\rho;j}} | P_1 \rangle
- \frac{1}{2}
{\mit\Pi}_{\sigma\tau} \frac{\partial}{\partial q_\tau}
\sum_{j = 1}^{\infty} \frac{1}{j}
\langle P_2 | {^V\!\tilde {\cal R}^2_{\rho;j}} | P_1 \rangle
\right\} .
\end{eqnarray}
Here we used a new convention
\begin{equation}
\tilde {\cal R}^2_{\rho;j}
= \frac{q_{\mu_1}}{q^2} \dots \frac{q_{\mu_j}}{q^2}
{\cal R}^2_{\rho; \mu_1 \dots \mu_j} ,
\end{equation}
and exploited an identity
\begin{eqnarray*}
\frac{\partial}{\partial \frac{q_\sigma}{q^2}}
= q^2 {\mit\Pi}_{\sigma\tau} \frac{\partial}{\partial q_\tau} .
\end{eqnarray*}

To evaluate the sums in (\ref{Fourier}) we substitute
(\ref{ChebyshevResumm}) in (\ref{LocalOperDD}) and replace
$x_\mu$ by $q_\mu/q^2$. This gives the local traceless matrix elements
$\langle P_2 | {^V\!\tilde {\cal R}^2_{\rho;j}} | P_1 \rangle$ in terms
of Chebyshev polynomials. The summation in Eq.\ (\ref{Fourier}) can
be now done making use of the generating function for the Chebyshev
polynomials $G_U (a, b) \equiv \sum_{j = 0}^{\infty} a^j U_j (b) =
(1 - 2 a b + a^2)^{- 1}$. To get a required $j$-dependent coefficient in
the series we do integration(s) of both sides with appropriate weight.
The sums required in the calculation of (\ref{Fourier}) read
\begin{eqnarray}
\label{ChebyshevSum}
&&\sum_{j = 0}^{\infty} \frac{a^{j + 1}}{(j + 1)} U_{j + 1} (b)
= \int_0^{a} \frac{da'}{a'}
\left( G_U (a', b) - 1 \right) , \\
&&\sum_{j = 1}^{\infty} \frac{a^{j + 1}}{j(j + 1)} U_{j + 1} (b)
= \int_0^a da' \int_0^{a'} \frac{da''}{\left( a'' \right)^2}
\left( G_U (a'', b) - 1 - 2 a'' b \right) .
\nonumber
\end{eqnarray}
Here $a = \ft12 \sqrt{{\cal P}^2/q^2}$ and $b = q \cdot
{\cal P}/\sqrt{q^2 {\cal P}^2}$.
This immediately leads to
\begin{eqnarray}
\label{FouTra1}
&&\sum_{j = 0}^{\infty}
\langle P_2 | {^V\!\tilde {\cal R}^2_{\rho;j}} | P_1 \rangle
=
q^2 {\mit\Pi}_{\rho\tau} \frac{\partial}{\partial q_\tau}
\int_{{\mit\Omega}} dy \, dz\, f (y, z) \\
&&\qquad\times
\left\{
\frac{1 - \sqrt{1 + {\cal M}^2}}{2 \sqrt{1 + {\cal M}^2}}
\ln\left(1 + \frac{1 - \sqrt{1 + {\cal M}^2}}{2{\mit\Xi}} \right)
-
\frac{1 + \sqrt{1 +{\cal M}^2}}{2 \sqrt{1 + {\cal M}^2}}
\ln\left(1 + \frac{1 + \sqrt{1 + {\cal M}^2}}{2{\mit\Xi}} \right)
\right\} , \nonumber\\
\label{FouTra2}
&&\sum_{j = 1}^{\infty} \frac{1}{j}
\langle P_2 | {^V\!\tilde {\cal R}^2_{\rho;j}} | P_1 \rangle
= - q^2 {\mit\Pi}_{\rho\tau} \frac{\partial}{\partial q_\tau}
\int_{{\mit\Omega}} dy \, dz\, f (y, z) \nonumber\\
&&\qquad \times
\Bigg\{ \frac{1}{{\mit\Xi}}
+ \frac{1 - \sqrt{1 + {\cal M}^2}}{2 \sqrt{1 + {\cal M}^2}}
\left( 1 + \frac{1 - \sqrt{1 + {\cal M}^2}}{2{\mit\Xi}} \right)
\ln\left(1 + \frac{1 - \sqrt{1 + {\cal M}^2}}{2{\mit\Xi}} \right) \nonumber\\
&&\qquad\qquad\ \ -
\frac{1 + \sqrt{1 + {\cal M}^2}}{2 \sqrt{1 + {\cal M}^2}}
\left( 1 + \frac{1 + \sqrt{1 + {\cal M}^2}}{2{\mit\Xi}} \right)
\ln\left(1 + \frac{1 + \sqrt{1 + {\cal M}^2}}{2{\mit\Xi}} \right)
\Bigg\} ,
\end{eqnarray}
where we introduced the conventions
\begin{equation}
{\mit\Xi} \equiv - \frac{q^2}{q \cdot {\cal P}} ,
\qquad
{\cal M}^2
\equiv - \frac{q^2 {\cal P}^2}{\left( q \cdot {\cal P} \right)^2} .
\end{equation}

Inserting our findings (\ref{FouTra1}) and (\ref{FouTra2}) into
Eq.\ (\ref{Fourier}) the next step is a mere differentiation w.r.t.\
$q_\mu$, which is done by the formula
\begin{equation}
\label{DiffQ}
q^2 {\mit\Pi}_{\mu\nu} \frac{\partial}{\partial q_\nu}
\tau \left( {\cal M}^2, {\mit\Xi} \right)
= \left\{
2 {\cal M}^2 \left( q_\mu + {\mit\Xi} {\cal P}_\mu \right)
\frac{\partial}{\partial{\cal M}^2}
- {\cal P}_\mu \frac{\partial}{\partial{\mit\Xi}^{-1}}
\right\}
\tau \left( {\cal M}^2, {\mit\Xi} \right)
\end{equation}
for a test function $\tau$. Finally, we contract the resulting equation
with the tensor $S_{\mu\nu;\rho\sigma}$ which leads to the hadronic tensor
\begin{equation}
\label{Tw2Amplitude}
T^2_{\mu \nu} = - \frac{1}{q^2} \int_{{\mit\Omega}} dy \, dz \, f (y, z)
\left\{
\left( q^2 g_{\mu\nu} - q_\mu q_\nu \right) {\cal F}_1
+
\left( {\cal P}_\mu + \frac{q_\mu}{{\mit\Xi}} \right)
\left( {\cal P}_\nu + \frac{q_\nu}{{\mit\Xi}} \right) {\cal F}_2
\right\} ,
\end{equation}
with mass-dependent coefficient functions
\begin{eqnarray}
\label{F1}
&&{\cal F}_1
=
\frac{4 {\mit\Xi} - {\cal M}^2 \left( 1 - {\mit\Xi} \right)}
{{\mit\Xi} \left[ 4 {\mit\Xi} (1 + {\mit\Xi}) - {\cal M}^2 \right]}
+ \frac{{\cal M}^2 \left( 2 {\mit\Xi} - {\cal M}^2 \right)}
{4 {\mit\Xi} \left( 1 + {\cal M}^2 \right)^{3/2}}
\ln \left(
\frac{1 - \sqrt{1 + {\cal M}^2} + 2 {\mit\Xi}}{1
+ \sqrt{1 + {\cal M}^2} + 2 {\mit\Xi}}
\right)
+ \left( {\mit\Xi} \to - {\mit\Xi} \right) ,
\nonumber\\
\label{F2}
&&{\cal F}_2
= \frac{ {\mit\Xi}
\left[ 4 {\mit\Xi} - {\cal M}^2 \left( 1 - {\mit\Xi} \right) \right]}
{(1 + {\cal M}^2) \left[ 4 {\mit\Xi} (1 + {\mit\Xi}) - {\cal M}^2 \right]}
+ \frac{ 3 {\mit\Xi} {\cal M}^2 \left( 2 {\mit\Xi} - {\cal M}^2 \right)}
{4 \left( 1 + {\cal M}^2 \right)^{5/2}}
\ln \left(
\frac{1 - \sqrt{1 + {\cal M}^2} + 2 {\mit\Xi}}{1
+ \sqrt{1 + {\cal M}^2} + 2 {\mit\Xi}}
\right) \\
&&\qquad\qquad\qquad\qquad\qquad\qquad\qquad
\qquad\qquad\qquad\qquad\qquad\qquad\qquad\qquad\quad \
+ \left( {\mit\Xi} \to - {\mit\Xi} \right) .
\nonumber
\end{eqnarray}
In order to get the right $i0$-prescription, so as to pick up a correct
sheet of the Riemann surface for the logarithm, we notice that in
(\ref{Integral}) we have to restore the suppressed Feynman prescription
as follows $q^2 \to q^2 + i 0$.

Let us comment on our result. As we observe, the leading order massless
(generalized) Callan--Gross-type relation (see \cite{BelMulKirSch00})
${\cal F}_2 = {\mit\Xi}^2 {\cal F}_1$ is violated by target mass
corrections. We checked that our result coincides in the forward limit
with the well-known ones \cite{GeoPol76}. Obviously, we have $q_\mu
T^2_{\mu \nu} = 0$ and therefore current conservation is fulfilled in
the forward case. Note that this is only the case in the sum of the
direct and the crossed amplitudes. However, the gauge invariance in
the off-forward kinematics is violated, i.e.\ $q_{2\mu} T^2_{\mu \nu}
\neq 0$ and, as it was previously studied in the case of massless
amplitudes, it is restored once higher twist corrections are accounted
for \cite{AniPirTer00,BelMul00,KivPolSchTer00,RadWei00}. This must also
persist to the case of traceless operators which are not projected
on the light cone.

Now let us generalize the formalism to a spin-$\ft12$ target. The matrix
elements $\langle P_2|{\cal R}^2_{\rho; j} |P_1\rangle$ are build now from
the vectors $P$, $\Delta$ and a set of (independent) Dirac bilinears,
characterizing the spin content of the target. For our purposes it is
convenient to use the following structures
\begin{equation}
\left( h_\mu, \tilde h_\mu \right)
= \bar U (P_2) \gamma_\mu \left( 1, \gamma_5 \right) U (P_1),
\qquad
\left( b, \tilde b \right)
= \bar U (P_2) \left( 1, \gamma_5 \right) U (P_1),
\end{equation}
for (vector, axial) sectors, respectively. Thus, we parametrize the matrix
element of e.g.\ vector operator according to
\begin{eqnarray}
\label{LorentzVscal}
\langle P_2|
{^V\!\!{\cal O}^{\rm sym}_{\rho; \mu_1 \dots \mu_j}}
| P_1 \rangle
\!\!\!&=&\!\!\!
h_{\{\rho} P_{\mu_1} \cdots P_{\mu_j\}} A_{j+1, j+1}
+ \cdots
+ h_{\{ \rho} \Delta_{\mu_1} \cdots \Delta_{\mu_j \}} A_{j+1, 1}
\\
\!\!\!&+&\!\!\!
\frac{b}{2M}
\left\{
P_{\{ \rho} P_{\mu_1} \cdots P_{\mu_j \}} B_{j+1, j+1}
+ \cdots
+ \Delta_{\{ \rho} \Delta_{\mu_1} \cdots \Delta_{\mu_j \}} B_{j+1, 0}
\right\} + \dots,
\nonumber
\end{eqnarray}
where again terms proportional to the metric tensor are not needed. The
non-local twist-two vector operator reads
\begin{eqnarray}
\langle P_2 | {^V\!{\cal R}^2_\rho} (x, - x) | P_1 \rangle
\!\!\!&=&\!\!\!
\int_{{\mit\Omega}} dy \, dz \,
\left\{
f_A (y, z) \, h \cdot \partial^{\cal P}
+
f_B (y, z) \, \frac{b}{2M} \, {\cal P} \cdot \partial^{\cal P}
\right\} \\
&&\qquad\qquad\qquad\times
\partial_\rho \sum_{j = 0}^{\infty} \frac{(-i)^{j}}{j!(j + 1)^2}
\left( \frac{x^2 {\cal P}^2}{4} \right)^{(j + 1)/2}
U_{j + 1} \left( \frac{x \cdot {\cal P}}{\sqrt{x^2 {\cal P}^2}} \right).
\nonumber
\end{eqnarray}
Here we have generated the factor of $x$ in $h\cdot x$ by a differentiation
w.r.t.\ ${\cal P}$. The same we did with the $b$ form factor for the purpose
of a uniform representation, as well as to have a cross check on our previous
calculations. Obviously, this step is not required at all and we can readily
borrow the results already found for a (pseudo) scalar target. The same
equation holds for axial operator with a trivial dressing of symbols with
tildes.

The reduced matrix elements $B_{j + 1, j + 1 - k}$ are represented in terms
of GPDs or DDs as in Eq.\ (\ref{BversusDD}), --- replace $f(y, z)$ by
$f_B (y, z)$. For the reduced matrix elements $A_{j + 1, j + 1 - k}$
we write
\begin{eqnarray}
\label{Def-GPD-mom-A}
A_{j, j - k}
\!\!\!&=&\!\!\! \frac{1}{k!}
\left. \frac{\partial^k}{\partial\eta^k} \right|_{\eta = 0}
\int_{-1}^{1} dx\; x^{j - 1} A(x,\eta)
= \left({j-1 \atop k}\right) \int_{{\mit\Omega}} dy \, dz \,
y^{j-k-1} z^{k} f_A(y,z) ,
\end{eqnarray}
where $j \geq 1$ and the index $k$ now varies in the intervals $0 \le k
\le j - 1$. There is an immediate consequence of the fact that $j$-th moment
of the function $A (x, \eta)$ is a polynomial of order $\eta^{j - 1}$ only.
Namely, translating it to the conventional Ji's distributions, $H = A + B$,
$E = - B$, we immediately find that
\begin{eqnarray*}
\frac{\partial}{\partial\eta^j} \int_{-1}^1 dx x^{j - 1}
\left\{ H (x, \eta) + E (x, \eta) \right\} = 0 .
\end{eqnarray*}
As a particular example we recall that for the $j = 2$ moment, which
corresponds to Ji's sum rule, the $\eta$-dependence drops off as it was
noted before \cite{Ji96}. For the axial channel the GPDs $\tilde A$ and
$\tilde B$ are identical to the conventional $\widetilde H$ and $\widetilde
E$ and this implies that the $j$-th moment of $\widetilde H$ ($\widetilde
E$) has the expansion to order $j - 1$ ($j$) in $\eta$.

The Fourier transform and the resummation are done in the same way as
before and result in
\begin{eqnarray}
{T^2_{\mu\nu}}
\!\!\!&=&\!\!\! - \frac{1}{q^2} \int_{{\mit\Omega}} dy \, dz \,
\Bigg\{
\left( \tilde f_A (y, z) \, \tilde h \cdot \partial^{\cal P}
+
\tilde f_B (y, z) \, \frac{\tilde b}{2M} \,
{\cal P} \cdot \partial^{\cal P}
\right)
i \epsilon_{\mu\nu\rho\sigma} q_\rho {\cal P}_\sigma
\ {\cal G}_1 \\
&&\qquad\qquad\quad\
+ \left( f_A (y, z) \, h \cdot \partial^{\cal P}
+
f_B (y, z) \, \frac{b}{2M} \, {\cal P} \cdot \partial^{\cal P}
\right) \nonumber\\
&&\quad\qquad\qquad\qquad\qquad\qquad\times
\left(
\left( q^2 g_{\mu\nu} - q_\mu q_\nu \right) {\cal F}_1
+
\left( {\cal P}_\mu + \frac{q_\mu}{{\mit\Xi}} \right)
\left( {\cal P}_\nu + \frac{q_\nu}{{\mit\Xi}} \right) {\cal F}_2
\right) \nonumber
\Bigg\},
\end{eqnarray}
with massive coefficient functions
\begin{eqnarray}
\label{Fn1}
{\cal F}_1
\!\!\!&=&\!\!\!
\frac{[2 {\mit\Xi} \left( 1 + 2 {\cal M}^2 \right) - {\cal M}^4]L_-}
{4 {\mit\Xi} \left( 1 + {\cal M}^2 \right)^{3/2}}
- \frac{L_+}{2 \left( 1 + {\cal M}^2 \right)}
- \frac{{\cal M}^2 \, L}{2 \left( 1 + {\cal M}^2 \right)^{3/2}}
+ \left( {\mit\Xi} \to - {\mit\Xi} \right) ,
\nonumber\\
\label{Fn2}
{\cal F}_2
\!\!\!&=&\!\!\!
\frac{{\mit\Xi}
[ 2 {\mit\Xi} \left( 1 + 4 {\cal M}^2 \right) - 3 {\cal M}^4 ] L_-}
{4 \left( 1 + {\cal M}^2 \right)^{5/2}}
- \frac{{\mit\Xi}^2 (1 - 2 {\cal M}^2) L_+}{2 \left( 1 + {\cal M}^2 \right)^2}
- \frac{3 {\mit\Xi}^2 {\cal M}^2 \, L}{2
\left( 1 + {\cal M}^2 \right)^{5/2}}
+ \left( {\mit\Xi} \to - {\mit\Xi} \right) , \nonumber\\
\label{Gn1}
{\cal G}_1
\!\!\!&=&\!\!\!
\frac{{\mit\Xi} \, L_-}
{2 \left( 1 + {\cal M}^2 \right)^{1/2}}
- \frac{{\mit\Xi} \, L_+}{2 \left( 1 + {\cal M}^2 \right)}
- \frac{{\mit\Xi} {\cal M}^2 \, L}{2 \left( 1 + {\cal M}^2 \right)^{3/2}}
- \left( {\mit\Xi} \to - {\mit\Xi} \right) ,
\end{eqnarray}
where
\begin{eqnarray*}
L_\pm
\!\!\!&\equiv&\!\!\!
\ln \frac{1 - \sqrt{1 + {\cal M}^2} + 2 {\mit\Xi}}{2 {\mit\Xi}}
\pm
\ln \frac{1 + \sqrt{1 + {\cal M}^2} + 2 {\mit\Xi}}{2 {\mit\Xi}}, \\
L
\!\!\!&\equiv&\!\!\!
{\rm Li}_2 \left( - \frac{1 - \sqrt{1 + {\cal M}^2}}{2 {\mit\Xi}} \right)
-
{\rm Li}_2 \left( - \frac{1 + \sqrt{1 + {\cal M}^2}}{2 {\mit\Xi}} \right) .
\end{eqnarray*}
Here ${\rm Li}_2$ is the Euler dilogarithm ${\rm Li}_2 (x) = - \int_0^x
\ft{d y}{y} \ln (1 - y)$.

As we mentioned before, the resummed mass corrections can not be expressed
in terms of GPDs (at least we did not succeed in doing this). However, every
term in the mass expansion, $M^2/q^2$, can be converted into the conventional
GPDs representation. Let us demonstrate it for the expanded Compton form
factor $F_1$ for a (pseudo) scalar target given in Eq.\ (\ref{F1}). To the
first non-trivial order, i.e.\ to ${\cal O} \left( M^4/q^4 \right)$ accuracy,
the latter reads in DD form
\begin{eqnarray}
\label{F1-expa}
F_1 \!\!\!
&=&\!\!\! \int_{{\mit\Omega}} dy \, dz \, f (y, z)
\Bigg\{
\Bigg(
C_1^{(0)}\! \left( {\mit\Xi}^{-1} - i 0 \right)
+
C_1^{(0)}\! \left( - {\mit\Xi}^{-1} - i 0 \right)
\Bigg) \nonumber\\
&&\qquad\qquad\qquad\qquad -
\frac{ M^2}{q^2} y^2 {\mit\Xi}^2
\Bigg(
C_1^{(1)}\! \left( {\mit\Xi}^{-1} - i 0 \right)
+
C_1^{(1)}\! \left( - {\mit\Xi}^{-1} - i 0 \right)
\Bigg)
\Bigg\},
\nonumber
\end{eqnarray}
where in our approximation, $M^2 \gg \Delta^2$, we set ${\cal M}^2 = - 4
\frac{M^2}{q^2} y^2 {\mit\Xi}^2$ and introduced the coefficient functions
$C_1^{(0)}\! \left( {\mit\Xi}^{-1} \right) = - \left( 1 + {\mit\Xi}
\right)^{-1}$ and $C_1^{(1)}\! \left( {\mit\Xi}^{-1} \right) = \left( 1
+ {\mit\Xi} \right)^{-2} + 2  \ln \left( \frac{{\mit\Xi}}{1 + {\mit\Xi}}
\right)$. To cast the DD into the usual GPD representations, which
are related by the equation $x \int_{{\mit\Omega}} dy \, dz \, f (y, z)
\delta (x - y - \eta z) \equiv B (x, \eta)$, we need the following general
result
\begin{equation}
\label{DDtoGPD1}
\int_{{\mit\Omega}} dy \, dz \,
\left\{
\!\!
\begin{array}{c}
z^n \\
y^n
\end{array}
\!\!
\right\}
\, {\cal F} \left( {\mit\Xi}^{-1} \right) f (y, z) \\
=
\int dx {\cal F} \left( \ft{x}{\xi} \right) \int dx' \,
 V^{(n)}_1 (x, x')
\left\{
\!\!
\begin{array}{c}
\left( \ft{\stackrel{\rightarrow}{\partial}}{\partial \eta} \right)^n \\
\prod_{k = 0}^{n - 1}
\left(
\hat d (x', \eta) - k
\right)
\end{array}
\!\!
\right\}
B (x', \eta) ,
\end{equation}
where we used a shorthand notation for the differential operator $\hat d
(x, \eta) = x \ft{\stackrel{\leftarrow}{\partial}}{\partial x} - \eta
\ft{\stackrel{\rightarrow}{\partial}}{\partial \eta}$ and the kernels
\begin{eqnarray}
V^{(n)}_1 (x, x')
=
\frac{(x' - x)^{n - 1}}{(n - 1)!} {\mit\Theta}^0_{11} (x, x - x'),
\quad \mbox{with} \quad
{\mit\Theta}^0_{11} (x, y)
= \frac{\theta (x) - \theta (y)}{x - y}.
\end{eqnarray}
Substitution of these expressions back into Eq.\ (\ref{F1-expa}) results into
\begin{eqnarray}
F_1
\!\!\! &=&\!\!\! \int_{-1}^1\!\! dx \int_{-1}^1 \!\! dx^\prime\;
\Bigg\{
\Bigg(
C_1^{(0)}\! \left( \ft{x}{\xi} - i 0 \right)
+
C_1^{(0)}\! \left( - \ft{x}{\xi} - i 0 \right)
\Bigg)
\delta (x - x^\prime)
\\
&&\qquad\qquad -
\frac{M^2}{q^2}
\Bigg(
C_1^{(1)}\! \left( \ft{x}{\xi} - i 0 \right)
+
C_1^{(1)}\! \left( - \ft{x}{\xi} - i 0 \right)
\Bigg)
V_1^{(2)} (x, x')  \left( \hat d (x', \eta) - 1 \right) \hat d (x', \eta)
\Bigg\}
B (x^\prime,\eta) , \nonumber
\end{eqnarray}
which has now the desired form. Completely analogous manipulations produce
the GPD representation for all other form factors.

\section{Conclusions.}
\label{Sec-Con}

In the present note, we have presented a formalism for the resummation of
target mass corrections stemming from the twist-two approximation to the
off-forward Compton scattering amplitude. The machinery is based on the
group-theoretical content of the procedure of trace subtraction from local
operators with definite symmetry property -- the operators with a well
defined twist are given by a harmonic projection of the latter. This allows
to resum target mass corrections to all orders in a compact form in terms
of DDs. However, once the amplitudes are expanded in $M^2/q^2$ series, they
can be cast in the form involving GPDs.

An important feature of the result is that the mass corrections enter with
additional powers of the scaling variable $\xi$. Thus, we expect that they
are suppressed in the small $\xi$ ($x_{\rm Bj}$) region. Since there still
might be a numerical enhancement due to coefficient functions, a numerical
model dependent study of the target mass effects to the deeply virtual
Compton scattering process for the kinematics of HERMES and Jefferson Lab
experiments is necessary.

In the twist-two approximation the current conservation is violated and must
be restored by taking into account contributions with higher geometrical
twists. To make the former manifest, one has to decompose those higher twist
operators in a linear independent set by means of the QCD equation of motion
\cite{BelMul00,KivPolSchTer00,RadWei00}. This procedure gives us a part
which is expressed in terms of total derivatives acting on operators with
a lower twist. Of course, the latter when combined together with the leading
prediction must render a manifestly gauge invariant result. However, the
trace subtraction and the application of the QCD equation of motion do not
commute and one can easily fail to reproduce an amplitude which respects
current conservation. This problem goes beyond the scope of the present
study and requires further investigations.

\vspace{1cm}

We would like to thank A.V. Radyushkin for a clarifying discussion of Ref.\
\cite{RadWei00}.

\end{document}